\documentclass[12pt]{article}
\usepackage{epsfig,graphicx}

\title{Reconstructing the cosmic evolution of the chemical elements}
\author{Anna Frebel}
\date{}

\begin{document}
\maketitle

\begin{abstract} %125-150words                                                                                                          
The chemical elements are created in nuclear fusion processes in the
hot and dense cores of stars. The energy generated through
nucleosynthesis allows stars to shine for billions of years. When
these stars explode as massive supernovae, the newly made elements are
expelled, chemically enriching the surrounding regions. Subsequent
generations of stars are formed from gas that is slightly more
element enriched than that from which previous stars formed. This
chemical evolution can be traced back to its beginning soon after the
Big Bang by studying the oldest and most metal-poor stars still
observable in the Milky Way today. Through chemical analysis, they
provide the only available tool for gaining information about the
nature of the short-lived first stars and their supernova explosions
more than thirteen billion years ago. These events set in motion the
transformation of the pristine universe into a rich cosmos of
chemically diverse planets, stars, and galaxies.

\end{abstract}

\vspace{1cm}

{\huge{O}}ne beautiful afternoon I went for a run along the river. I
was breathing plenty of fresh air, my face was all flushed, and I felt
my heart pounding and blood flowing through my body. As air was
filling my lungs, I was reminded of Carl Sagan's saying: ``We are all
made from star stuff.'' Indeed we are. When quenching my thirst with
water, I was consuming hydrogen and oxygen in the form of
H$_{2}$O. When breathing, I had been taking in air made from nitrogen,
oxygen, and tiny traces of other elements such as argon and neon. The
red liquid of life owes its color to iron, which is embedded in our
hemoglobin. But these elements do not just circulate within our
carbon-based bodies: before they became part of humans, each of these
atoms was created in a grand cosmic cycle called chemical evolution
that took place long before biological evolution led to life on Earth.

Most of the universe's iron, for example, is the end result of a
binary star system in which one star acquires enough material from
its companion that it reaches a critical mass and erupts in a huge
thermonuclear explosion, forging new elements in the process. On the
other hand, the hydrogen atoms that make up water are probably
nearly fourteen billion years old and were created as part of the Big
Bang. And all the carbon upon which life as we know it is based was
synthesized in evolved stars near the end of their lives.

The fact that all elements except hydrogen, helium, and lithium are
made in stars and their subsequent explosions has only been known for
less than sixty years. A seminal paper from 1957, often referred to as
``B2FH'' following the initials of the authors, provided the first
comprehensive summary of ``the synthesis of the elements in
stars.''$^1$ This came after decades of work directed at finding the
energy source of stars. With the elucidation of how and where the
chemical elements are forged in stars came the realization that there
is a chemical evolution in the universe, causing a net increase in the
amount of elements over time. Most important, this model provided
observationally testable support for the Big Bang theory and the
theory of a time-dependent chemical evolution of the universe.

While the nature of chemical evolution of galaxies is now well
established, many details of the complex circle of nucleosynthesis in
stars, later chemical enrichment of interstellar gas, and subsequent
star formation remain poorly understood and thus continue to be
subject to ongoing research. Many questions center around what the
exact abundance yields of individual supernova explosions may be, as
well as how the nature of the exploding stars themselves and the
astrophysical environment influences nucleosynthesis and the
production of the elements throughout the periodic table. Because old
stars that formed in the early phases of chemical evolution can help
with this quest, we will start the tale of the origin of the elements
from the very beginning of the universe.

{\Huge {I}}mmediately after the Big Bang 13.8 billion years ago, there
was a time without stars and galaxies. The hot gas left over from the
Big Bang had to cool enough before the first cosmic objects were
able to form. This process took a few hundred million years, but
eventually the very first stars lit up the universe. The universe at
that time was made from just hydrogen and helium: heavier elements did
not exist yet. As a consequence of a variety of gas chemistry and
cooling processes that govern star formation, the first stars are
thought to have been rather massive. Recent computations suggest
these behemoths may have had up to one hundred times the mass of the
sun$^2$. In comparison, most stars today are low-mass stars with less
than one solar mass.

Stars are powered by the nuclear fusion taking place in their cores;
it is the energy source that sustains their enormous luminosities. In
the first and by far the longest burning phase, hydrogen is fused into
helium. At about ten million degrees F, four protons (or hydrogen
nuclei) are fused together in a series of nuclear reactions to make a
helium nucleus. Subsequent burning stages, which occur only in the
last ten percent of stars' lives, result in three helium nuclei
(``$\alpha$-particles'') being converted to beryllium, which then
captures another $\alpha$-particle to become carbon in the so-called
triple-$\alpha$ process.

After that, through additional $\alpha$-particle captures, carbon
nuclei are converted to oxygen; through yet more nucleosynthesis
processes, all elements in the periodic table up to iron are built
up. The fusion of lighter elements into heavier ones results in a
conversion of a small amount of mass into energy. For example, a
helium nucleus is 0.7 percent lighter than four individual hydrogen
nuclei. It is this mass difference that, as described by $E = mc^{2}$,
fuels the star and sustains its luminosity for long periods of
time. However, once the star has created an iron core at its center,
nucleosynthesis stops. No more energy can be gained by fusing iron
into even heavier nuclei: the star's energy source has ceased for
good. As a consequence, the star can no longer maintain equilibrium
and begins to collapse due to its own gravity. As a result of the huge
pressures, the iron core is converted into an extremely dense neutron
star. The collapsing mass of the star bounces off the hard neutron
star and leads to a gigantic supernova, leaving the neutron star
behind. This was also the fate of the massive very first stars. To
sustain their great luminosities, massive stars (those with more than
ten solar masses) require large amounts of nuclear
energy. Consequently, they burned through the hydrogen and
subsequently created heavier-element fuel much more quickly than stars
with lower masses, therefore limiting their lifetimes to just a few
million years.

During the explosion of a star, all the newly created elements are
released into the surrounding gas. The death of the first stars
marked an important milestone in the evolution of the universe: it was
not pristine anymore, but ``polluted'' with carbon, oxygen, nitrogen,
iron, and other elements. Thus, over time, the universe became
more and more enriched in the elements heavier than hydrogen and
helium, which are collectively called ``metals'' by astronomers. In
contrast, the very first stars were the only ones that formed from
completely metal-free gas. All stars in subsequent generations
would then form from gas clouds that contained some metals provided
by at least one previous generation of stars exploding as supernovae.

The sudden existence of metals in the early universe following the
death of the first stars changed the conditions for subsequent star
formation. Gas clouds can cool down more efficiently when metals or
dust made from metals are present, leading to the collapse of smaller
clouds, and thus the formation of smaller stars. Lower-mass stars like
the sun could therefore form for the first time. The first low-mass
stars (those with 60 to 80 percent of the mass of the sun) have long
lifetimes of fifteen to twenty billion years due to their sparse
consumption of the nuclear fuel in their cores. Born soon after the
Big Bang as secondor third-generation stars, they are still shining
today. Many of these ancient survivors are suspected to be hiding in
our Milky Way galaxy and, indeed, astronomers have discovered dozens
of them over the past three decades. What makes these extremely rare
objects so valuable is that they preserve in their atmospheres
information about the chemical composition of their birth cloud, which
existed soon after the Big Bang. Hence, studying their chemical
composition allows astronomers to reconstruct the early era of their
births.

In the earliest stages of the universe's development, massive stars
exploding as supernovae dominated the production of iron in the
universe. However, this changed after about a billion years. Through
the existence of the first lower-mass stars with longer lifetimes, a
different pathway for iron production emerged. At the end of their
long lives, low-mass stars turn into compact white dwarf remnants. If
a star and a white dwarf are in a binary system and enough mass is
transferred from the star to the white dwarf, the latter will undergo
a thermonuclear explosion. Given the dominance of low-mass stars in
the universe today, iron is thus mainly produced by this process
rather than by exploding massive stars, as was exclusively the case in
the early universe.

After about nine billion years of this chemical evolution, driven by
different types of stars at different times, our sun, together with
its planets, finally formed. Its birth gas had been enriched by
perhaps a thousand generations of stars and supernova explosions. That
evolution pro vided the gas with enough metals to enable the formation
of planets–something that may not have been possible much earlier on
in the universe. Consequently, when astronomers look for extrasolar
planets, they focus their search on stars that are close in age to or
younger than the sun.

{\Huge{T}}hrough spectroscopic observations, astronomers can determine
which elements are present in a star's outer layers and what their
respective abundances are. Spectroscopy is a technique in which
starlight is split up into its components, just as sunlight is split
when we see a rainbow. The different elements (hydrogen, helium, and
metals) in the star's atmosphere absorb light at very specific colors,
or wavelengths. When carrying out high-resolution spectroscopy, the
starlight is significantly stretched out over all visible wavelengths
to enable the detection of even very weak absorption lines left behind
by all the elements in the stars. The existence and strength of
absorption lines corresponding to specific elements are measured and
analyzed with computer programs that reconstruct the stellar
atmosphere. This way, astronomers can calculate how many atoms of a
given element are present in the star.

High-resolution spectroscopy, especially for fainter stars, requires
the largest telescopes that observe the visible wavelength
range. Telescopes like the Magellan-Clay Telescope at Las Campanas
Observatory, located in Chile's Atacama desert, are equipped with
high-resolution spectrographs. Thanks to its large 6.5-meter-diameter
mirror, the Magellan Telescope is capable of collecting enough light
from faint stars to enable high-resolution spectroscopic
measurements. Chemical analysis then shows how much of each type of
metal is present in a star, which indicates the star's formation
time. So-called metal-poor stars are assumed to be old because they
formed from gas enriched with only a trace amount of heavy elements,
created by the first few stellar generations after the Big Bang.$^3$ In
contrast, ``metal-rich'' stars like the sun must have formed at a much
later time when the universe was significantly enriched with metals by
many stellar generations.

Our study of early star formation and chemical evolution relies on our
ability to measure stars' metallicity, or metal content. The main
indicator used to determine stellar metallicity is iron abundance,
which, with few exceptions, reflects a star's overall metallicity
fairly well. Absorption lines of iron (Fe) can be found throughout
stellar spectra, often covering large wavelength ranges from 350 to
900 nanometers, which makes measuring iron abundance relatively
straightforward. The iron abundance of a star is given as [Fe/H],
which is used in the logarithm of the ratio of iron atoms to hydrogen
atoms in comparison to that of the sun. The formal definition reads
$\mbox{[Fe/H]} = \log_{10}(N_{\rm Fe}/N_{\rm H})_{\star}
-\log_{10}(N_{\rm Fe}/N_{\rm H})_{\odot}$ with N being the number of Fe
and H atoms, respectively, and $\star$ and $\odot$ representing the star being
evaluated and the sun, respectively. The consequence of this
logarithmic definition is that metal-poor stars will have negative
[Fe/H] values, as those stars have a lower concentration of Fe atoms
than the sun. Stars containing higher concentrations of metals than
the sun will show a positive [Fe/H] value.

\begin{figure}[!t]
\begin{center}
%\vspace{5cm}
\includegraphics[width=11.7cm,clip=true,bbllx=64,bblly=420,bburx=537,bbury=654]{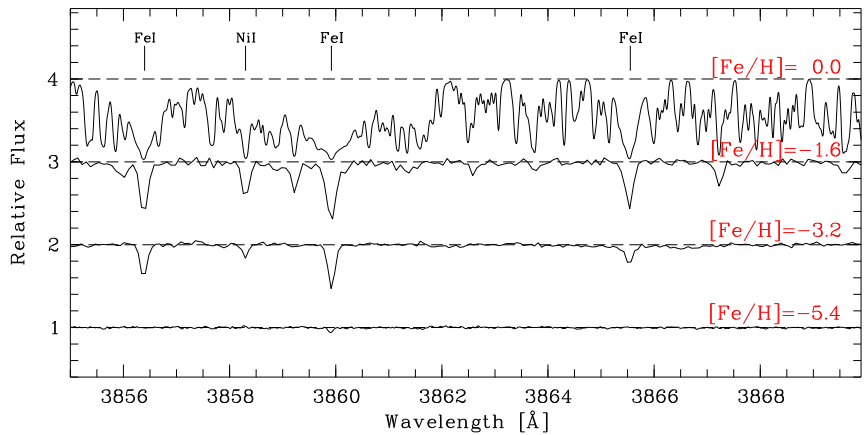}
  \caption{\label{spec_comp_plot} Spectral comparison of the Sun with
    three metal-poor stars with different metallicities indicating the
    course of chemical evolution from early universe (bottom) to the
    Sun's birth $\sim4.5$ billion years ago (top).  Source: Anna
    Frebel, ``Stellar Archaeology: Exploring the Universe with
    Metal-Poor Stars,'' Astronomische Nachrichten 331 (2010): 474.}
\end{center}
\end{figure}

To illustrate the difference between younger metal-rich and older
metal-poor stars, Figure~\ref{spec_comp_plot} shows spectra of the sun and three
metal-poor stars. Their decreasing metallicities are listed. The
corresponding number of absorption lines detectable in the spectra
decreases with increasing metaldeficiency. In star HE~1327$-$2326 
(bottom spectrum), only very few metal absorption lines are left to
observe. Their weakness is such that determining their metallicity
requires extremely high quality data that can only be obtained with
large telescopes.$^4$

If one wishes to identify the oldest stars, the task is to find stars
with the lowest metallicities and thus the earliest formation
times. It is those stars that allow astronomers to look back in time
and reconstruct the formation and evolution of the chemical elements
and the involved nucleosynthesis processes that created them. While
very distant galaxies are often used for observational studies of
galaxy formation and cosmology, metal-poor stars are the local
equivalent of the distant universe and thus the object of
``near-field'' cosmology. Both approaches to cosmology complement
each other in providing detailed information and observational
constraints that push us toward understanding the onset of star and
galaxy formation in the early universe some thirteen billion years
ago.  Metal-poor stars are, however, the only tool we have available
to learn about the nature of the first stars and their supernova
explosions. Our study of these stars therefore provides unique
constraints on various theoretical concepts regarding the physical
and chemical nature of the early universe.

Past sky surveys for metal-poor stars have shown that these ancient
objects can be systematically identified in a three-step process that
involves the selection of candidates from the survey data and
subsequent follow-up of the best targets with mediumand
high-resolution spectroscopy.$^5$ This technique has identified large
numbers of metal-poor stars on the outskirts of the Milky Way, the
so-called halo of the galaxy. Work done over the last few decades has
shown that stars with low metallicity are much fewer in number
compared to more metal-rich stars, reflecting not only the chemical
evolution of the universe but also the overwhelming number of stars
that have formed since its early stages.

The most metal-deficient stars, in particular, are extremely rare and
difficult to find. Only about fifty stars are known to have
metallicities of $\mbox{[Fe/H]} < -3.5$, which corresponds to
$\sim$1/3000th of the solar metallicity. Of those, only six have
$\mbox{[Fe/H]} < -4.0$ or $ <$1/10,000th of the solar value. The
current record holder is the star SMSS~0313$-$6708, with $\mbox{[Fe/H]}
< -7.0$. No iron lines could be detected, so only an upper limit on the iron
abundance could be determined, which corresponds to less than
$\sim$1/10,000,000th that of the sun. The next most iron-poor star,
HE~1327$-$2326, has an iron abundance of $\mbox{[Fe/H]} =-5.4$
($\sim$1/250,000th of the solar iron abundance). This translates into
an actual iron mass of just 1 percent of the iron mass present in the
Earth's core. This is a very small amount, considering that the star
is approximately 300,000 times more massive than the Earth and about
one million times larger in size. It also reveals that in the early
universe, iron and other elements were rare commodities.

Thus, the few stars with $\mbox{[Fe/H]} < -4.0$ have opened a new and
unique observational window to the time shortly after the Big Bang
when only the very first stars had enriched the universe. They are
frequently employed to constrain theoretical studies about the
formation of the first stars, element production and chemical
evolution, and supernova yields. The elemental-abundance patterns
(chemical abundances as a function of atomic number of the respective
element) of these stars appear to be highly individual, but in fact
can be successfully reproduced by scenarios in which a massive first
supernova explosion provided the elements to the gas cloud from which
the observed object later formed. In fact, the most metal-poor stars
all display the ``fingerprint'' of one single massive first supernova,
which allows astronomers to ascertain the mechanisms and details of
the supernova itself and the nature of the long-extinct progenitor
star.

{\Huge{W}}ith the exception of hydrogen and helium, all elements up to
iron are created through nuclear fusion during lifetimes of stars. But
these elements (with atomic number $Z < 30$) make up less than one third
of the periodic table. So where do the other elements with higher
atomic masses, such as silver ($Z = 47$) and gold ($Z = 79$), or more
exotic rare earth elements, such as lanthanum ($Z = 57$) and europium ($Z
= 63$), come from?

The study of metal-poor stars has greatly advanced our understanding
of this topic. As we now know from nuclear physics, elements heavier
than iron are created not through fusion processes but through
neutron-capture by seed nuclei (for example, iron nuclei). In an
astrophysical environment that provides a constant flux of neutrons,
heavy elements can thus be built up. Such conditions are thought to
occur during certain kinds of supernova explosions in which a strong
neutron flux develops above the newly formed central neutron star. For
example, if iron nuclei are extremely rapidly bombarded with many
neutrons before the nuclei $\beta$-decay, their nuclei capture more
neutrons, creating heavy, neutron-rich, and unstable isotopes. Once
these have $\beta$-decayed to stability, new and heavier elements
remain.$^6$ $\beta$-decay is a spontaneous decay of one element into
another through the conversion of a neutron into a proton accompanied
by the emission of an electron and a neutrino. Due to the rapid bombardment, this
process is called the r-process. About half of all stable isotopes of
elements heavier than zinc are made this way.

The other half of the isotopes of heavy elements are created in the
so-called sprocess, where a slower neutron bombardment (over a longer
timescale than the $\beta$-decay process) leads to the successive
buildup of heavy elements. This process occurs in the pulsing outer
shells of evolved red giant stars with masses of less than eight solar
masses and metallicities of $\mbox{[Fe/H]} > -3.0$ (indicating that
the star formed from gas already containing a small amount of iron
atoms that could function as seed nuclei). Through stellar winds,
these elements are eventually released into the surrounding gas.

In 1995, a low-mass, metal-poor  star, CS 22892-052, was
discovered to possess a very high abundance of numerous
neutron-capture elements (including the rare earth elements) compared
to lighter elements such as iron. Indeed, the star has a metallicity
of $\mbox{[Fe/H]} \sim -3.0$ (or $\sim$1/1000th of the solar iron
abundance), but the neutron-capture material is about forty times more
abundant. The various neutroncapture elements detected in this star
are likely the result of an r-process event that took place prior to
the star's birth. When the star formed, it inherited the chemical
signature of this particular nucleosynthesis event. For the
4.5-billion-year-young sun, which formed from gas enriched by many
generations of stars, it is possible to infer how much of each
observed element may have been produced by rprocess events prior to
the sun's formation. The resulting solar r-process pattern can be
compared to that of other, more metal-poor stars. A comparison between
the sun and the metal-poor CS 22892-052, for example, revealed that
both stars have the exact same relative pattern of neutroncapture
abundances (see Figure~2). It appears that at any time and place in
the universe, the r-process creates its heavy elements in the exact
same ratios, indicating that the r-process is a universal
process. Since most neutron-capture elements are too heavy to be
created and studied in accelerator laboratories on Earth, this has
been an important empirical finding based on stellar astronomy.

\begin{figure}[!t]
\begin{center}
%\vspace{}
\includegraphics[width=12.0cm,bbllx=27, bblly=205,
   bburx=555, bbury=690]{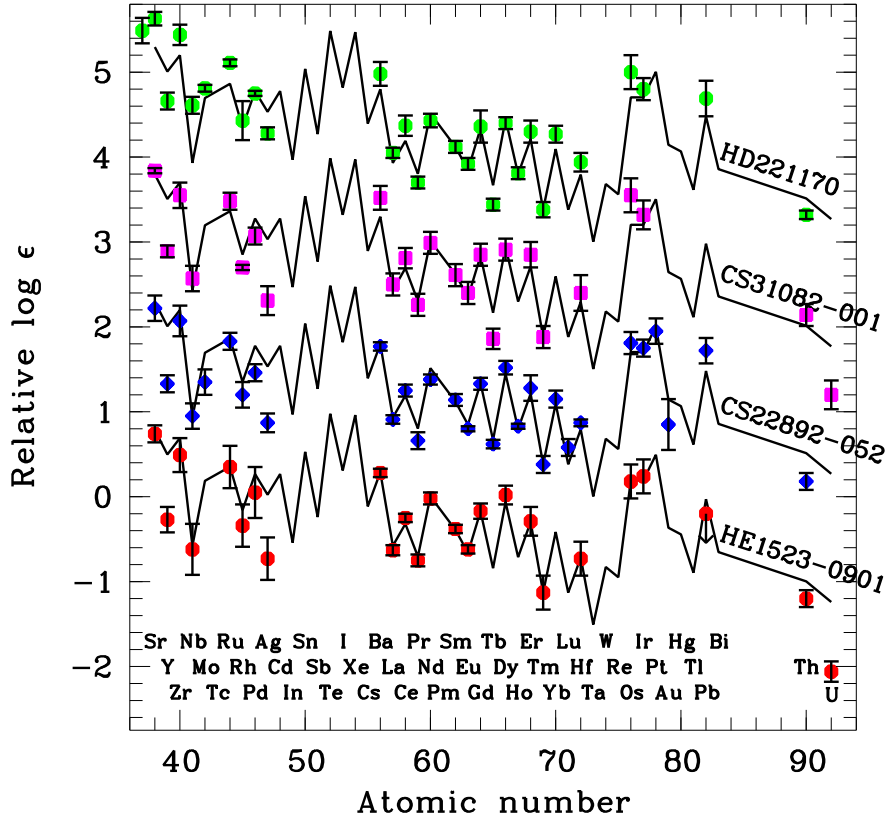}
  \caption{\label{rproc_plot} Elemental abundances in four metal-poor
    stars with a relative overabundance of $r$-process elements
    (various colored symbols) compared with the scaled solar r-process
    pattern (solid line). All patterns are arbitrarily offset to allow
    a visual comparison. Note the remarkable agreement of the
    metal-poor star pattern and that of the Sun for elements heavier
    than barium ($Z\ge56$). Source: Anna Frebel and John E. Norris,
    ``Metal-Poor Stars and the Chemical Enrichment of the Universe,''
    in Planets, Stars and Stellar Systems vol. 5, ed. Gerard Gilmore
    (Amsterdam: Springer, 2012), doi:10.1007/978-94-007-5612-0\_3.}
\end{center}
\end{figure}

{\Huge{T}}he elements produced in the r-process include thorium and
uranium, which are very long-lived radioactive isotopes:
$^{232}$Th has a half-life of 14 billion years while $^{238}$U has a
half-life of 4.5 billion years (they are thus decaying very slowly and
are near-stable on Earth). Measuring thorium and uranium abundance in
metal-poor stars whose birth gas cloud was enriched by only one or few
supernova events enables astronomers to carry out cosmo-chronometry:
dating the oldest stars with a method analogous to dating
archaeological finds through radiocarbon analysis. In the latter
technique, the initial ratio of $^{12}$C (the typical stable form of
carbon) to $^{14}$C (an unstable isotope) must be estimated and then
compared to the ratio at the time of discovery. Cosmo-chronometry
requires astronomers to know the initial amount of the heaviest
elements, which were presumably produced together in a massive
supernova explosion (obtaining such information is extremely
challenging, but detailed calculations of r-process nucleosynthesis
have yielded estimates). The estimated initial ratios of unstable and
stable rare earth elements (such as thorium to europium, uranium to
osmium, and thorium to uranium) can then be compared with the
currently observed ratios, and the degree of decay of the unstable
isotopes thus provides the age of the star.

While thorium is often detectable, uranium poses a great
challenge. Only one extremely weak absorption line of uranium is
available in the optical spectrum, making its detection difficult, if
not impossible, in most cases. In an ideal scenario, both radioactive
elements are detected so that many ratios of the thorium and/or
uranium abundance to those of stable rare earth elements can be
compared to model predictions for the yields of the r-process
event. Indeed, several r-process metal-poor stars with metallicities
of roughly 1/1000th of the solar value were found to be about 14
billion years old. These include HE~1523$-$0901, which is only the
third metal-poor star in which uranium can reliably be
detected. Moreover, HE~1523$-$0901 can be dated with seven different
``cosmic clocks''; that is, abundance ratios containing either thorium
or uranium and different rare earth elements. The average age obtained
through this analysis is 13.2 billion years; this is consistent with
the universe's age of 13.8 billion years, which has been deduced from
observations of the cosmic background radiation interpreted with the
latest cosmological models. Unfortunately, the range of uncertainty
with respect to stellar age is often several billion
years. Regardless, cosmo-chronometry confirms that HE~1523$-$0901 and
all other metal-poor stars are ancient and formed soon after the Big
Bang during the early phases of chemical evolution.

Through individual age measurements, metal-poor r-process stars
provide an independent lower limit for the age of the universe. This
makes them vital probes for near-field cosmology. At the same time,
given their rich inventory of very heavy, exotic nuclei, these stars
also closely connect astrophysics and nuclear physics by acting as a
``cosmic lab`` for both fields of study.

{\Huge{R}}ecent searches for metal-poor stars have not only focused on
the old stellar halo but also on dwarf satellite galaxies orbiting the
Milky Way. The ultra-faint dwarf galaxies–whose total luminosities
range from 1000 to 100,000 solar luminosities, making them the dimmest
galaxies known–appear to contain almost exclusively metal-poor
stars. These systems ran out of gas for additional star formation
billions of year ago. Chemical evolution and star formation ceased as
a result, and when we observe these systems, we can only see the
leftover low-mass stars that are still shining today. They, too, tell
us the story of nucleosynthesis and enrichment in the early stages of
the universe.$^7$ In fact, there are recent indications that these
systems are nearly as old as the universe itself: some of them may be
among the first galaxies that formed after the Big Bang. Studying
these stars thus offers another chance to reconstruct the initial
events of element creation within the first stars and their violent
explosions, and the subsequent incorporation of this material into
next-generation stars. Moreover, the existence of such old satellites
may shed light on the existence of metal-poor stars in the halo of the
Milky Way. Predating our own galaxy, these halo stars must have come
from somewhere; perhaps they originated from dwarf galaxies when
analogous systems were gobbled up by the Milky Way during its assembly
process.$^8$

Topics like these inspire astronomers to collect additional
information about the nature and structure of the galaxy. To
chemically characterize the galactic halo in detail, including its
streams, substructures, and satellites, wide-angle surveys with large
volumes are needed. The Australian SkyMapper Telescope is already
mapping the Southern sky. It is optimized for stellar work and is
delivering new metal-poor star candidates for which highresolution
spectroscopy will be required. The Chinese lamost spectroscopic survey
is providing numerous metal-poor candidates in the Northern
hemisphere. Studying ever-fainter stars further out in the deep halo
of the Milky Way and in faraway dwarf galaxies may become a reality
with the light-collecting power of the next generation of optical
telescopes, including the Giant Magellan Telescope, the Thirty Meter
Telescope, and the European Extremely Large Telescope. These
telescopes are currently scheduled for completion around 2020. At this
point, only the Giant Magellan Telescope is scheduled to be equipped
with the high-resolution spectrograph necessary to study metal-poor
stars. Further, gaia, an astrometric space mission led by the European
Space Agency (ESA) that was launched in late 2013, will obtain
high-precision astrometry for one billion stars in the galaxy, along
with the physical parameters and the chemical composition of many of
them. Together, these new data will revolutionize our understanding
of the origin, evolution, structure, and dynamics of the Milky Way.

All of these new observations will be accompanied by an increased
theoretical understanding of the first stars and galaxies, supernova
nucleosynthesis, and the mixing of metals into gas clouds in the early
universe, as well as cosmic chemical evolution. New generations of
sophisticated cosmological simulations of galaxy fomation and
evolution will enable a direct investigation of chemical evolution (in
a first-galaxy simulation, for example). Being able to trace the metal
production and corresponding spatial distributions will allow
astronomers to compare the results with abundance measurements of
metal-poor stars in the Milky Way's satellite dwarf galaxies. This way,
studying nucleosynthesis and the products of chemical evolution will
reveal whether any of the ultra-faint dwarf galaxies are surviving
first galaxies and whether the metal-poor galactic halo was assembled
from early analogs of today's dwarf satellites billions of years ago.

\section*{References}

[1] Burbidge, E. Margaret, Burbidge, Geoffrey R., Fowler, William
A. \& Hoyle, Fred ``Synthesis of the Elements in Stars'', Reviews of
Modern Physics, 29, 547 (1957)

\noindent
[2] Abel, Tom, Bryan, Greg~L. \& Norman, Michael~L. ``The Formation of
      the First Star in the Universe'', Science, 295, 93 (2002)

\noindent
[3] Frebel, Anna \& Norris, John~E. ``Metal-Poor Stars and the Chemical
Enrichment of the Universe`` to appear in Vol. 5 of textbook
``Planets, Stars and Stellar Systems'', by Springer, in 2013;
currently in arXiv1102.1748 (2011)

\noindent
[4] Frebel, Anna ``Stellar archaeology: Exploring the Universe with
metal-poor stars'', Astronomische Nachrichten, 331, 474 (2010)
  
\noindent
[5] Beers, Timothy~C. \& Christlieb, Norbert ``The Discovery and
    Analysis of Very Metal-Poor Stars in The Galaxy'', Annual Review
    of Astronomy \& Astrophysics, 43, 531 (2005)

\noindent
[6] Sneden, Christopher, Cowan, John~J. \& Gallino, Roberto
``Neutron-Capture Elements in the Early Galaxy'', Annual Review of
Astronomy \& Astrophysics, 46, 241-288 (2008)

\noindent
[7] Frebel, Anna \& Bromm, Volker ``Precious fossils of the infant
universe'', Physics Today, 65, 4, 49 (2012)

\noindent
[8] Frebel, Anna \& Bromm, Volker ``Chemical Signatures of the First
Galaxies: Criteria for One-shot Enrichment'', Astrophysical Journal,
759, 115 (2012)

\end{document}